\documentclass
[preprint,pre,nobibnotes,nofootinbib,notitlepage,10pt,twocolumn,showpacs,superscriptaddress]{revtex4}%
\usepackage{amsfonts}
\usepackage{amsmath}
\usepackage{amssymb}
\usepackage{graphicx}%
\begin{document}
\title{Alternative model of the Antonov problem}
\author{L. Velazquez}
\email{luisberis@geo.upr.edu.cu}
\affiliation{Departamento de F\'{\i}sica, Universidad de Pinar del R\'{\i}o, Mart\'{\i}
270, Esq. 27 de Noviembre, Pinar del R\'{\i}o, Cuba.}
\author{F. Guzm\'{a}n}
\email{guzman@info.isctn.edu.cu}
\affiliation{Departamento de F\'{\i}sica Nuclear, Instituto Superior de Ciencias y
Tecnolog\'{\i}a Nucleares, Carlos III y Luaces, Plaza, La Habana, Cuba.}
\date{\today}

\begin{abstract}
Astrophysical systems will never be in a real Thermodynamic equilibrium: they
undergo an evaporation process due to the fact that the gravity is not able to
confine the particles. Ordinarily, this difficulty is overcome by enclosing
the system in a rigid container which avoids the evaporation. We proposed an
energetic prescription which is able to confine the particles, leading in this
way to an alternative version of the Antonov isothermal model which unifies
the well-known isothermal and polytropic profiles. Besides of the main
features of the isothermal sphere model: the existence of the gravitational
collapse and the energetic region with a negative specific heat, this
alternative model has the advantage that the system size naturally appears as
a consequence of the particles evaporation.

\end{abstract}
\maketitle

\section{Introduction}

Thermodynamical properties of selfgravitating systems are very different from
the ones exhibited by the traditional systems. They are typical
\textit{nonextensive systems} since they are nonhomogeneous and the total
energy is not extensive, which is a consequence of the long-range character of
gravitational interaction. They also exhibit energetic regions with a
\textit{negative specific heat}, which persist even in the thermodynamic limit
\cite{pad,kies,ruelle,lind,antonov,Lynden,Lynden1,thirring}. That is the
reason why the Gibbs canonical ensemble is non applicable to the description
of selfgravitating systems, since this ensemble is not able to access to those
macroscopic states possesing a negative heat capacity. At first glance, the
selfgravitating systems could only be described by using the microcanonical ensemble.

Gravitation is not able to confine the particles: it is always possible that
some of them have the sufficient energy for escaping out from the system, so
that, the selfgravitating systems always undergo an evaporation process.
Therefore, they will never be in a real thermodynamic equilibrium. This
difficulty is usually avoided by enclosing the system in a rigid container
\cite{antonov,Lynden,Lynden1}, which could be justified when the evaporation
rate is small and certain kind of quasistationary state might be reached.
There are some approaches which have taken into account this kind of
regularization of the long-range singularity of the Newtonian potential in a
microcanonical framework \cite{gross,de vega}.

The use of the rigid container can be conveniently substituted by imposing the
following energetic prescription:%

\begin{equation}
\frac{1}{2m}p^{2}+m\phi\left(  \mathbf{r}\right)  \leq\epsilon_{S}%
<0,\label{renormalization}%
\end{equation}
where $\frac{1}{2m}p^{2}$ is the kinetic energy of a particle of mass $m$ and
$m\phi\left(  \mathbf{r}\right)  $ its gravitational potential energy at the
point $\mathbf{r}$, being $\epsilon_{S}$ an energy cutoff which is determined
from the existence of certain tidal forces. All those particles satisfying the
condition (\ref{renormalization}) are confined by the gravity, on contrary,
they will be able to scape out from the system if they do not lose their
excessive energy. This regularization procedure is characteristic of the
Michie-King model for globular clusters \cite{bin}, which supposes that those
stars that gain suficient velocity through encounters are able to escape or
are removed by tidal forces. The main motivation of such regularization scheme
relies on the consideration of evaporation effects, and therefore, this
regularization procedure is more realistic than the box regularization (the
use of the rigid container). The aim of this paper is to develop an
alternative model to the standard isothermal sphere model of Antonov
\cite{antonov} based on the consideration of this energetic prescription
starting from microcanonical basis.

\section{Statistical description}

Let us consider the \textit{N}-body selfgravitating Hamiltonian system:%

\begin{equation}
H_{N}=T_{N}+U_{N}=\underset{k=1}{\overset{N}{\sum}}\frac{1}{2m}\mathbf{p}%
_{k}^{2}-\overset{N}{\underset{j>k=1}{\sum}}\frac{Gm^{2}}{\left\vert
\mathbf{r}_{j}-\mathbf{r}_{k}\right\vert }.
\end{equation}
Taking into consideration the above regularization prescription
(\ref{renormalization}), the admissible stages of this system are those in
which the kinetic energy of a given particle satisfies the condition:%

\begin{equation}
\frac{1}{2m}\mathbf{p}_{k}^{2}<u_{k}=m\left[  \phi_{S}-\phi\left(
\mathbf{r}_{k}\right)  \right]  \text{,} \label{renpart}%
\end{equation}
being $\phi\left(  \mathbf{r}_{k}\right)  $ the gravitational potential at the
point $\mathbf{r}_{k}$ where the \textit{k}-th particle is located:%

\begin{equation}
\phi\left(  \mathbf{r}_{k}\right)  =-\overset{N}{\underset{j\neq k}{\sum}%
}\frac{Gm^{2}}{\left\vert \mathbf{r}_{k}-\mathbf{r}_{j}\right\vert },
\label{pot discreto}%
\end{equation}
where we introduce the tidal potential $\phi_{S}\,\ $($\epsilon_{S}=m\phi_{S}$).

\begin{figure}
[h]
\begin{center}
\includegraphics[
height=2.4526in,
width=2.6558in
]%
{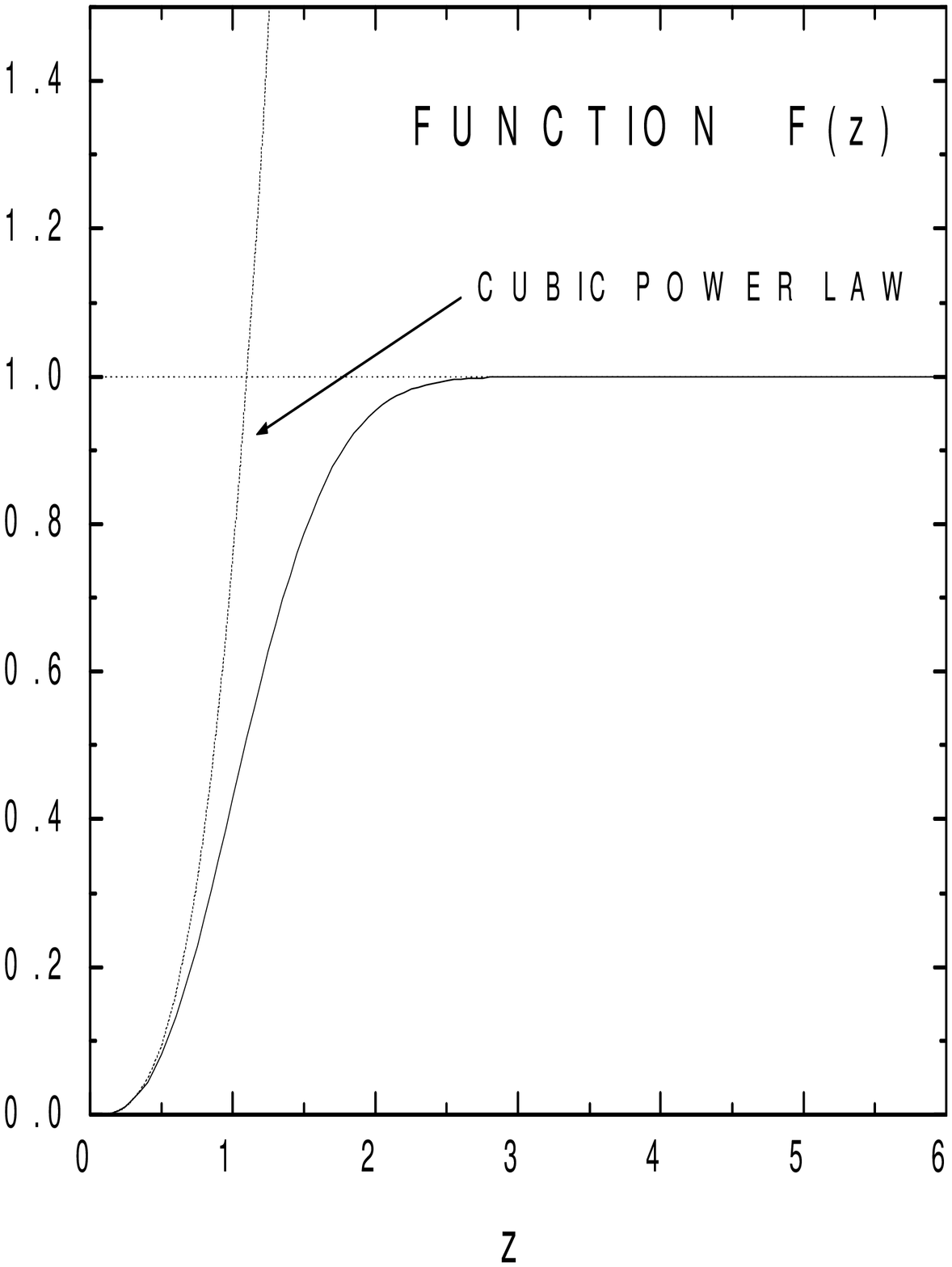}%
\caption{Representation of the function $F\left(  z\right)  $. Its symptotic
behaviors are also shown.}%
\label{fig1}%
\end{center}
\end{figure}

The regularized microcanonical accessible volume $W_{R}$ is given by:%

\[
W_{R}=\frac{1}{N!}\int_{X_{R}}\delta\left[  E-H_{N}\right]  dX,
\]
where $X_{R}$ is a subspace of the \textit{N}-body phase-space where the
condition (\ref{renpart}) takes place, being $dX=d^{3N}Rd^{3N}P/(2\pi
\hbar)^{3N}=\underset{k}{\prod}\frac{d^{3}\mathbf{r}_{k}d^{3}\mathbf{p}_{k}%
}{\left(  2\pi\hbar\right)  ^{3}}$ the volume element. The spacial coordinates
should be also regularized in order to avoid the short-range divergenge of the
Newtonian potential, which will be performed below by using the mean field
approximation. This integral is rewritten by using the Fourier representation
of the delta function as follows:%

\begin{equation}
W_{R}=\underset{-\infty}{\overset{+\infty}{\int}}\frac{dk}{2\pi}\exp\left(
zE\right)  \mathcal{Z}_{R}\left(  z,N\right)  ,
\end{equation}
where $\mathcal{Z}_{R}\left(  z,N\right)  $:%

\begin{equation}
\mathcal{Z}_{R}\left(  z,N\right)  =\frac{1}{N!}\int_{X_{R}}\exp\left(
-zH_{N}\right)  dX,\label{can partition}%
\end{equation}
is the canonical partition function with complex argument $z=\beta+ik\,$, with
$\beta\in\Re$. Integration by $d^{3N}P$ yields:%

\begin{equation}
\frac{1}{N!}\int_{\Re^{3N}}d^{3N}R\left(  \frac{m}{2\pi\hbar^{2}z}\right)
^{\frac{3}{2}N}\exp\left[  -zU_{N}\left(  R\right)  +\chi\left(  R;z\right)
\right]  ,\label{partition discreta}%
\end{equation}
being $\chi\left(  R;z\right)  $:%

\begin{equation}
\chi\left(  R;z\right)  =\underset{k=1}{\overset{N}{\sum}}\ln F\left[
\sqrt{zu_{k}}\right]  .
\end{equation}
where $F\left(  z\right)  $ is defined by:%

\begin{equation}
F\left(  z\right)  =\operatorname{erf}\left(  z\right)  -\frac{2}{\sqrt{\pi}%
}z\exp\left(  -z^{2}\right)  \text{.}%
\end{equation}
and shown in the figure (\ref{fig1}). The asymptotic behaviors of $F\left(
z\right)  $ are given by:
\begin{equation}
F\left(  z\right)  =\left\{
\begin{array}
[c]{cc}%
\frac{4}{3\sqrt{\pi}}z^{3}+O\left(  z^{3}\right)   & \text{when }%
z\rightarrow0\\
\sim1 & z\geq2.5
\end{array}
\right.  \text{.}\label{serie}%
\end{equation}
%

%TCIMACRO{\FRAME{fhFU}{2.6558in}{2.4526in}{0pt}{\Qcb{Representation of the
%function $F\left(  z\right)  $. Its symptotic behaviors are also shown.}%
%}{\Qlb{fig1}}{fig1.eps}{\special{ language "Scientific Word";
%type "GRAPHIC";  display "USEDEF";  valid_file "F";  width 2.6558in;
%height 2.4526in;  depth 0pt;  original-width 6.5613in;
%original-height 8.7623in;  cropleft "0";  croptop "1";  cropright "1";
%cropbottom "0";  filename 'Fig1.eps';file-properties "XNPEU";}}}%
%BeginExpansion
%EndExpansion

We are interested in describing the large $N$ limit. This aim can be carried
out by using following procedure: we partition the physical space in cells
$\left\{  c_{\alpha}\right\}  $ being $\mathbf{r}_{\alpha}$ the positions of
their centers. We denoted by $n_{\alpha}=n\left(  \mathbf{r}_{\alpha}\right)
$ the number of particles inside the cell at the position $\mathbf{r}_{\alpha
}$. Introducing the density $\rho\left(  \mathbf{r}_{\alpha}\right)  =n\left(
\mathbf{r}_{\alpha}\right)  /v_{\alpha}$, being $v_{\alpha}$ the volume of the
cell $c_{\alpha}$, the functions $U_{N}\left(  R\right)  $ and $\chi\left(
R;z\right)  $ are rewritten by using a \textit{mean field approximation} as follows:%

\begin{equation}
U_{N}\left(  R\right)  \longrightarrow U\left[  \rho,\phi\right]  =\int
_{\Re^{3}}\frac{1}{2}m\rho\left(  \mathbf{r}\right)  \phi\left(
\mathbf{r}\right)  d^{3}\mathbf{r,}%
\end{equation}

\begin{align}
\chi\left(  R;z\right)   &  \longrightarrow\chi\left[  \rho,\phi;z\right]
\nonumber\\
&  =\int_{\Re^{3}}\rho\left(  \mathbf{r}\right)  \ln F\left(  \sqrt{zm\left[
\phi_{S}-\phi\left(  \mathbf{r}\right)  \right]  }\right)  d^{3}\mathbf{r,}%
\end{align}
being $\phi\left(  \mathbf{r}\right)  $ the Newtonian potential for a given
$\rho$ profile:%

\begin{equation}
\phi\left(  \mathbf{r}\right)  =\mathcal{G}\left[  \rho\right]  =-\int
_{\Re^{3}}\frac{Gm\rho\left(  \mathbf{r}_{1}\right)  d^{3}\mathbf{r}_{1}%
}{\left\vert \mathbf{r}-\mathbf{r}_{1}\right\vert }, \label{green}%
\end{equation}
which is the Green solution of the problem:%

\begin{equation}
\Delta\phi=4\pi G\rho.
\end{equation}

It is easy to understand that this mean field approximation acts as a partial
regularization procedure for the short-range singularity of the Newtonian
potential. The microscopic fluctuations of the Newtonian potential are
disregarded in this approximation, since this field is considered constant
inside the volume of each cell. Thus, the gravity effects on the microscopic
picture of each cell are reduced to the truncation of the velocity
distribution of the particles. This regularization is partial since it does
not avoid the gravitational collapse \cite{antonov}.

Using the partition of the physical space in cells, the integration by
$d^{3N}R$ can be approximately given by:%

\begin{equation}
\frac{1}{N!}\int_{\Re^{3N}}d^{3N}R\simeq\underset{\left\{  n_{\alpha}\right\}
}{\sum}\delta_{D}\left(  N-\underset{\alpha}{\sum}n_{\alpha}\right)
\underset{\alpha}{\prod}\frac{v_{\alpha}^{n_{\alpha}}}{n_{\alpha}!},
\end{equation}
where:%

\begin{equation}
\delta_{D}\left(  k\right)  =\left\{
\begin{tabular}
[c]{ll}%
$1,$ & if $k=0$\\
$0$ & otherwise
\end{tabular}
\ \ \right.  \text{ and }\underset{\left\{  n_{\alpha}\right\}  }{\sum}%
\equiv\underset{n_{1}}{\sum}\underset{n_{2}}{\sum}\cdots.
\end{equation}

The following factor can be rephrased in the mean field approximation as follows:%

\begin{equation}
\left(  \frac{m}{2\pi\hbar^{2}z}\right)  ^{\frac{3}{2}N}\underset{\alpha
}{\prod}\frac{v_{\alpha}^{n_{\alpha}}}{n_{\alpha}!}\rightarrow e^{-p_{f}%
\left[  \rho,z\right]  },
\end{equation}
being%

\begin{equation}
p_{f}\left[  \rho,z\right]  =\int_{\Re^{3}}d^{3}\mathbf{r}\text{ }\rho\left(
\mathbf{r}\right)  \left[  \ln\rho\left(  \mathbf{r}\right)  -1+\frac{3}{2}%
\ln\left(  \frac{2\pi\hbar^{2}z}{m}\right)  \right]  ,
\end{equation}
where the Stirling formula $\ln n!\simeq n\ln n-n$ was used. Finally, we
rephrase the summation in the mean field approximation as follows:
\begin{equation}
\underset{\left\{  n_{\alpha}\right\}  }{\sum}\delta_{D}\left(  N-\underset
{\alpha}{\sum}n_{\alpha}\right)  \longrightarrow\int\mathcal{D}\rho\left(
\mathbf{r}\right)  \delta\left[  N-\int_{\Re^{3}}d^{3}\mathbf{r}\text{ }%
\rho\left(  \mathbf{r}\right)  \right]  .
\end{equation}

Taking into consideration all approximations introduced above, the canonical
partition function (\ref{can partition}) is rewritten as:%

\[
\mathcal{Z}_{c}\left[  z,N\right]  =\int\mathcal{D}\rho\left(  \mathbf{r}%
\right)  \delta\left[  N-N\left[  \rho\right]  \right]  \times
\]

\begin{equation}
\times\exp\left\{  -p_{f}\left[  \rho,z\right]  -zU\left[  \rho,\phi\left[
\rho\right]  \right]  +\chi\left[  \rho,\phi\left[  \rho\right]  ;z\right]
\right\}  ,\label{mf ca}%
\end{equation}
being $N\left[  \rho\right]  $ the particles number functional:%

\begin{equation}
N\left[  \rho\right]  =\int_{\Re^{3}}d^{3}\mathbf{r}\text{ }\rho\left(
\mathbf{r}\right)  .
\end{equation}
In order to avoid the complicate $\rho$ dependence of $\phi\left[
\rho\right]  $ in (\ref{mf ca}), we introduce the identity:%

\begin{equation}
\int\mathcal{D}\phi\left(  \mathbf{r}\right)  \delta\left\{  \phi\left(
\mathbf{r}\right)  -\mathcal{G}\left[  \rho\right]  \right\}  =1,
\end{equation}
in the functional integral (\ref{mf ca}):%

\[
\int\mathcal{D}\rho\left(  \mathbf{r}\right)  \mathcal{D}\phi\left(
\mathbf{r}\right)  \delta\left\{  \phi\left(  \mathbf{r}\right)
-\mathcal{G}\left[  \rho\right]  \right\}  \delta\left(  N-N\left[
\rho\right]  \right)  \times
\]

\begin{equation}
\times\exp\left\{  -p_{f}\left[  \rho;z\right]  -zU\left[  \rho,\phi\right]
+\chi\left[  \rho,\phi;z\right]  \right\}  .
\end{equation}
The delta functions are also conveniently rewritten by using their Fourier representation:%

\begin{equation}
\delta\left\{  \phi\left(  \mathbf{r}\right)  -\mathcal{G}\left[  \rho\right]
\right\}  \sim\int\mathcal{D}h\left(  \mathbf{r}\right)  \exp\left[  \int
_{\Re^{3}}d^{3}\mathbf{r}J\left(  \mathbf{r}\right)  \left\{  \phi\left(
\mathbf{r}\right)  -\mathcal{G}\left[  \rho\right]  \right\}  \right]  ,
\label{delta fi}%
\end{equation}
as well as%

\begin{equation}
\delta\left(  N-N\left[  \rho\right]  \right)  =\underset{-\infty}%
{\overset{+\infty}{\int}}\frac{dq}{2\pi}\exp\left[  z_{1}\left(  N-N\left[
\rho\right]  \right)  \right]  ,
\end{equation}
being $z_{1}=\mu+iq$ with $\mu\in\Re$; $J\left(  \mathbf{r}\right)  =j\left(
\mathbf{r}\right)  +ih\left(  \mathbf{r}\right)  $, a complex function with
$j\left(  \mathbf{r}\right)  $ and $h\left(  \mathbf{r}\right)  \in\Re$. Thus,
the canonical partition function $\mathcal{Z}_{c}\left[  z,N\right]  $ is
finally expressed as:%

\begin{equation}
\underset{-\infty}{\overset{+\infty}{\int}}\frac{dq}{2\pi}\int\mathcal{D}%
\rho\left(  \mathbf{r}\right)  \mathcal{D}\phi\left(  \mathbf{r}\right)
\mathcal{D}h\left(  \mathbf{r}\right)  \exp\left\{  z_{1}N-H\left[  \rho
,\phi;z,z_{1},J\right]  \right\}  .
\end{equation}
The functional $H\left[  \rho,\phi;z,z_{1},J\right]  $ is given by:%

\begin{equation}
=p_{f}\left[  \rho;z\right]  +zU\left[  \rho,\phi\right]  -\chi\left[
\rho,\phi;z\right]  +z_{1}N\left[  \rho\right]  +K\left[  \phi,\rho,J\right]
,\label{h functional}%
\end{equation}
being $K\left[  \phi,\rho,J\right]  $ the exponential argument of the
expression (\ref{delta fi}).

The reader may check that when $N$ is scaled as $N\rightarrow\alpha N$, and
the following quantities are scaled as follows:%

\begin{equation}%
\begin{tabular}
[c]{l}%
$\rho\rightarrow\alpha^{2}\rho,\text{ }\mathbf{r}\rightarrow\alpha^{-\frac
{1}{3}}\mathbf{r,}\text{ }\phi\rightarrow\alpha^{\frac{4}{3}}\phi,$\\
$z\rightarrow\alpha^{-\frac{4}{3}}z,~z_{1}\rightarrow z_{1},~J\rightarrow
\alpha^{\frac{2}{3}}J,$%
\end{tabular}
\text{ } \label{sca b}%
\end{equation}
all terms in (\ref{h functional}) scale proportional to $\alpha$, and therefore:%

\begin{equation}
H\left[  \rho,\phi;z,z_{1},J\right]  \rightarrow\alpha H\left[  \rho
,\phi;z,z_{1},J\right]  .\label{sca b2}%
\end{equation}
The thermodynamic limit is carry out tending $\alpha$ to the infinity,
$\alpha\rightarrow\infty$. Thus, we can estimate $\mathcal{Z}_{c}\left[
z,N\right]  $ for $N$ large by using the \textit{steepest decent method}.
$\,$The Planck\ potential $\mathcal{P}\left(  \beta,N\right)  =-\ln
\mathcal{Z}_{c}\left[  \beta,N\right]  $ is thus obtained as follows:%

\begin{equation}
\mathcal{P}\left[  \beta,N\right]  \simeq-\underset{\rho,\text{ }\phi}{\max
}\left\{  \underset{\text{ }\mu,\text{ }j}{\min}\left[  \mu N-H\left[
\rho,\phi;\beta,\mu,j\right]  \right]  \right\}  .
\end{equation}
where the stationary conditions:%

\begin{equation}
\frac{\delta H}{\delta\rho}=\frac{\delta H}{\delta\phi}=\frac{\delta H}{\delta
j}=\frac{\delta H}{\delta\mu}-N=0,
\end{equation}
lead to the following relations:%

\begin{align}
\rho &  =\left(  \frac{m}{2\pi\hbar^{2}\beta}\right)  ^{\frac{3}{2}}%
\exp\left[  -\mu-\frac{1}{2}\beta m\phi+\mathcal{G}\left[  j\right]  \right]
\times\nonumber\\
&  \times F\left[  \sqrt{\beta m\left[  \phi_{S}-\phi\right]  }\right]  ,
\label{e1}%
\end{align}

\begin{equation}
j=-\frac{1}{2}\beta m\rho+\rho\partial_{\phi}\ln F\left[  \sqrt{\beta m\left[
\phi_{S}-\phi\right]  }\right]  , \label{e2}%
\end{equation}

\begin{equation}
\phi=\mathcal{G}\left[  \rho\right]  \text{ and }N=N\left[  \rho\right]  ,
\label{e3}%
\end{equation}

The relations (\ref{e1}) and (\ref{e2}) define the state equation; and the
relations (\ref{e3}) establish the Newtonian potential $\phi$ for a given
$\rho$ profile as well as the normalization constrain for the number of
particles. The relation (\ref{e1}) is conveniently rewritten by using
(\ref{e2}) as follows:%

\begin{equation}
\rho=N\left(  \beta,\mu\right)  \exp\left[  \Phi+C\right]  F\left(
\Phi^{\frac{1}{2}}\right)  , \label{RF}%
\end{equation}
where $\Phi=\beta m\left[  \phi_{S}-\phi\right]  $ and $N\left(  \beta
,\mu\right)  =\left(  \frac{m}{2\pi\hbar^{2}\beta}\right)  ^{\frac{3}{2}}%
\exp\left(  -\mu-\beta m\phi_{S}\right)  $, being $C$ a new function which is
given by:%

\begin{equation}
C=-\beta m\mathcal{G}\left[  \rho\partial_{\Phi}\ln F\left(  \Phi^{\frac{1}%
{2}}\right)  \right]  . \label{CF}%
\end{equation}
It is not difficult to show that the Planck potential is given by:%

\begin{equation}
\mathcal{P}\left(  \beta,N\right)  =-\left(  1+\mu+\frac{1}{2}\beta m\phi
_{S}\right)  N+\int_{\Re^{3}}d^{3}\mathbf{r}\text{ }\rho\left(  \frac{1}%
{2}\Phi+C\right)  . \label{planck FP}%
\end{equation}

The Boltzmann entropy can be estimated by using the steepest decent method as follows:%

\begin{equation}
S_{B}\left(  E,N\right)  \simeq\underset{\beta}{\min}\left[  \beta
E-\mathcal{P}\left[  \beta,N\right]  \right]  ,\label{entropy stp}%
\end{equation}
being $E\left[  \beta,\rho,\phi\right]  $ the energy functional:%

\begin{equation}
E\left[  \beta,\rho,\phi\right]  =\int_{\Re^{3}}d^{3}\mathbf{r}\text{ }\left[
3-\Phi\partial_{\Phi}\ln F\left(  \Phi^{\frac{1}{2}}\right)  \right]  \frac
{1}{2\beta}\rho+\frac{1}{2}m\rho\phi. \label{energy tot}%
\end{equation}
It can be easily seen that the quantity:%

\begin{equation}
\epsilon\left(  \beta,\Phi\right)  =\left[  3-\Phi\partial_{\Phi}\ln F\left(
\Phi^{\frac{1}{2}}\right)  \right]  \frac{1}{2\beta},
\label{ener per particle}%
\end{equation}
represents the \textit{kinetic energy per particle} at a given point of the
selfgravitating system. Note that $\rho$ vanishes when $\Phi$ tends to zero in
(\ref{RF}), so that, the particle distribution has been \textit{regularized}.
This state equation differs from the one obtained in the isothermal model by
the presence of the truncation function $F\left(  \Phi^{\frac{1}{2}}\right)  $
as well as the \textit{driving function} $C$. Although $C$ naturally appears
in our derivation, its existence is closely related with the modification
provoked in the microscopic picture of the system by the evaporation. An
example of this affirmation is the deviation of the Maxwell distribution along
the system, which can be noted in the kinetic energy per particle
$\epsilon\left(  \beta,\Phi\right)  $ (\ref{ener per particle}). A naive
energy truncation of the \textit{Maxwell-Boltzmann distribution}:%

\begin{equation}
\omega_{MB}=C_{0}\exp\left[  -\beta\left(  \frac{1}{2m}p^{2}+m\phi\right)
\right]  .
\end{equation}
leads to the state function:%

\begin{equation}
\rho\sim\exp\left[  \Phi\right]  F\left(  \Phi^{\frac{1}{2}}\right)  ,
\end{equation}
but here the driving function $C$ does not appear.

According to the scaling laws (\ref{sca b}), this astrophysical model obeys to
the following thermodynamic limit:%

\begin{equation}
N\rightarrow\infty,\text{ keeping constant }\frac{E}{N^{\frac{7}{3}}}\text{
and }LN^{\frac{1}{3}}, \label{THL}%
\end{equation}
being $L$ the characteristic system size. This thermodynamic limit was
obtained in ref.\cite{chava} for the self-gravitating fermions model. The
reader may surprise of the $N$-dependence of the characteristic system size
$L$. However, there is nothing strange in this behavior since the
selfgravitating gas is constituted by \textit{punctual} particles, and
therefore, this model system can be reduced to a point in the thermodynamic
limit $N\rightarrow\infty$.

It is easy to understand that this is an asymptotical behavior which
disappears when the particles size or/and the relativistic effects are taken
into account. It means that it should exist a superior limit of $N$ in which
these \textit{N}-dependences of the energy and system size become invalid, and
therefore, the selfgravitating nonrelativistic gas model is inapplicable for
describing the thermodynamical properties of such massive astrophysical systems.

It is interesting to analyze the asymptotic behavior of the particles density
state function $\rho\left(  \Phi;\beta,\mu\right)  $. According to the
asymptotic behaviors (\ref{serie}) of the function $F\left(  z\right)  $, the
state equation (\ref{RF}) becomes in the \textit{isothermal distribution} when
$\Phi>2.5$\thinspace:%

\begin{equation}
\rho\propto\exp\left[  \Phi\right]  ,
\end{equation}
which is characteristic of the inner regions of the system (the core). At
local level, the particles velocities obey to a Maxwell distribution, where
the kinetic energy per particle (\ref{ener per particle}) is given by
$3/2\beta$. When we move\ from the inner regions towards the outer ones, the
kinetic energy per particle decreases until zero, being this approximately
given by $\epsilon\simeq3\Phi/5\beta$. There the particles velocities obey to
an \textit{uniform distribution}, and so, the state equation in the halo obey
to a \textit{polytropic distribution}:%

\begin{equation}
\rho\propto\frac{4}{3\sqrt{\pi}}\Phi^{\frac{3}{2}}. \label{poly}%
\end{equation}

\begin{figure}
[h]
\begin{center}
\includegraphics[
height=2.4526in,
width=2.5823in
]%
{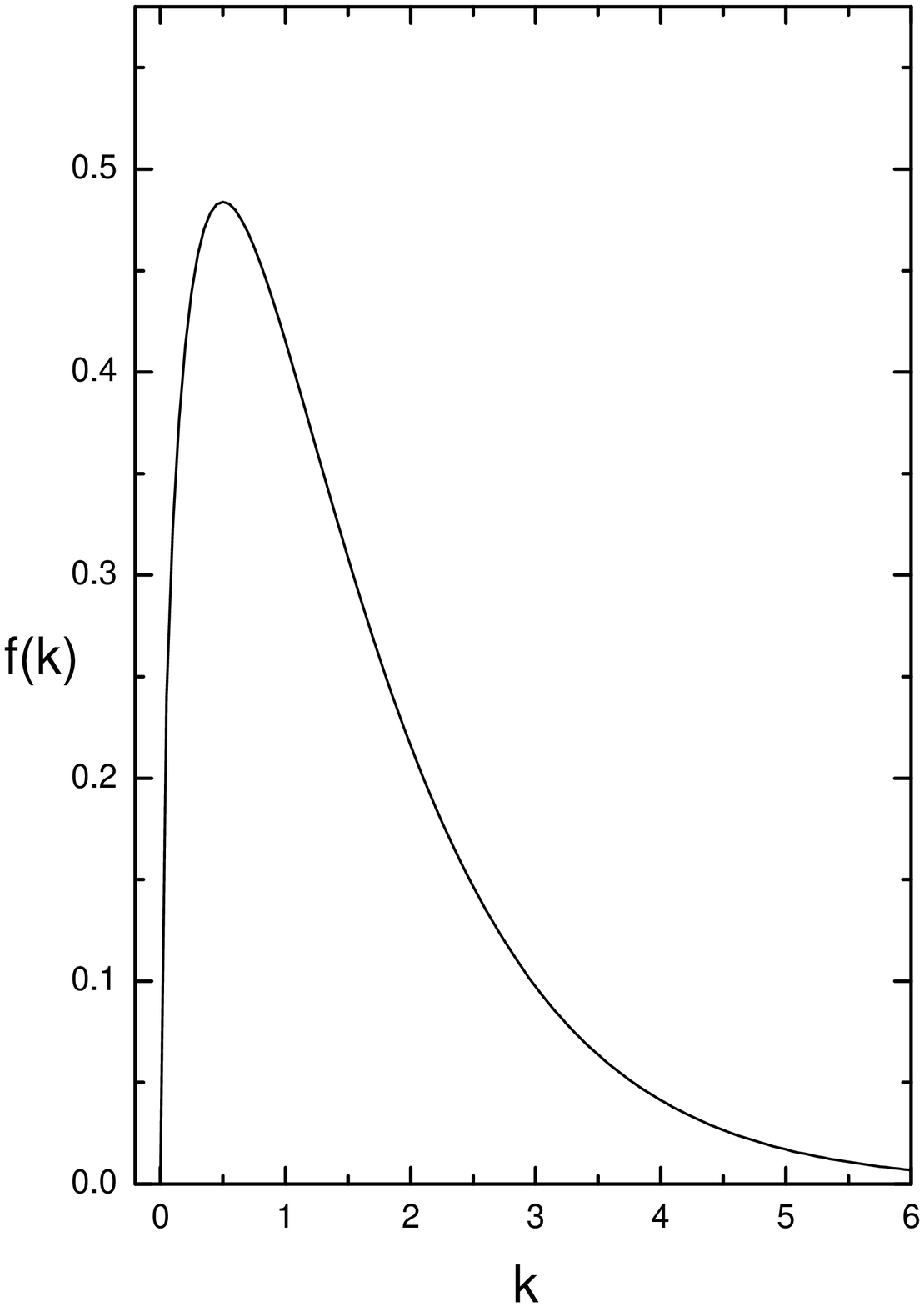}%
\caption{Maxwell distribution for the kinetic energy. The character of the
state function for the particles density $\rho$ depends on the localization of
the energy cuttof $\Phi$.}%
\label{max}%
\end{center}
\end{figure}

All these behaviors appear as consequence of the high energy cutoff
(\ref{renpart}) in the Maxwell distribution for the kinetic energy $f\left(
k\right)  =2\pi^{-\frac{1}{2}}k^{\frac{1}{2}}\exp\left(  -k\right)  $
($k=\beta\mathbf{p}^{2}/2m$ ), which is shown in the figure (\ref{max}). Note
that the cutoff value of $k$ is equal to $\Phi$. In the inner regions where
$\Phi>2.5$, this energy cutoff is unimportant because of it is located in the
tail of the distribution. However, it modifies considerably the character of
the state function for $\rho$ in the halo due to the fact that it is located
before the pick. The particles density $\rho$ will obey to the polytropic
dependence (\ref{poly}) throughout the whole system volume when $\Phi$ is also
small in the inner regions.%

%TCIMACRO{\FRAME{fhFU}{2.5823in}{2.4526in}{0pt}{\Qcb{Maxwell distribution for
%the kinetic energy. The character of the state function for the particles
%density $\rho$ depends on the localization of the energy cuttof $\Phi$.}%
%}{\Qlb{max}}{max.eps}{\special{ language "Scientific Word";  type "GRAPHIC";
%display "USEDEF";  valid_file "F";  width 2.5823in;  height 2.4526in;
%depth 0pt;  original-width 6.3503in;  original-height 9.0001in;
%cropleft "0";  croptop "1";  cropright "1";  cropbottom "0";
%filename 'max.eps';file-properties "XNPEU";}}}%
%BeginExpansion
%EndExpansion

Thus, the consideration of the energetic prescription (\ref{renormalization})
always leads to the existence of a polytropic halo, where the system core can
be isothermal or polytropic, according to the values of the function $\Phi$ in
the inner regions of the system. Due to the general properties of the
spherical solutions with a polytropic profile, the particles density will
\textit{vanish} at a finite radio $R_{tidal}$, which can be identify with the
characteristic size of the system. This radio is related with the tidal
potential $\phi_{S}$ throughout the relation:%

\begin{equation}
\phi_{S}=-\frac{GM}{R_{tidal}},
\end{equation}
being $M$ the system total mass. Thus, the characteristic size of the system
is \textit{determined} by the tidal interactions.

\section{Numerical study}

In order to perform a numerical study of the equations obtained in the
previous section, we express the energy in units of $E_{0}=GM^{2}/R_{tidal}$
and the lenght in units of the system size $R_{tidal}$, and the mass in unit
of $M$. It is not difficult to show that the functions $\Phi$ and $C$ obey to
the following structure equations:%

\begin{equation}
\Delta_{r}\Phi=-4\pi F_{1}\left(  \Psi,\Phi\right)  ,~\Delta_{r}\Psi=-4\pi
F_{2}\left(  \Psi,\Phi\right)  \text{\ ,} \label{structure equation}%
\end{equation}
where $\Delta_{r}u=r^{-2}\partial_{r}\left(  r^{2}\partial_{r}u\right)  $ is
the radial part of the Laplace operator, being the functions $F_{1}\left(
\Psi,\Phi\right)  $ and $F_{2}\left(  \Psi,\Phi\right)  $ defined by:%

\begin{align}
F_{1}\left(  \Psi,\Phi\right)   &  =\exp\left(  \Psi+\Phi\right)  F\left(
\Phi^{\frac{1}{2}}\right)  ,\nonumber\\
F_{2}\left(  \Psi,\Phi\right)   &  =\frac{2}{\sqrt{\pi}}\exp\left(
\Psi\right)  \Phi^{\frac{1}{2}}.\label{FF}%
\end{align}
In the expressions above the function $\Psi=C+\ln\mathcal{K}\left(  \beta
,\mu\right)  $, with $\mathcal{K}\left(  \beta,\mu\right)  =\beta^{-\frac
{1}{2}}\exp\left(  -\mu+\beta\right)  $. The solutions of the nonlinear system
(\ref{structure equation}) satisfy the following conditions at the surface:%

\begin{equation}
\Phi\left(  1\right)  =0,\text{ }\Phi^{\prime}\left(  1\right)  =-\beta
,~C\left(  1\right)  =-C^{\prime}\left(  1\right)
\end{equation}
which are derived from the boundary conditions of Green solutions (\ref{green}).

The solution can be obtained from the imposition of the following boundary
conditions at the origin:%

\begin{align}
\Phi\left(  0\right)   &  =\Phi_{0}>0\text{,}\\
\Psi\left(  0\right)   &  =\Psi\left(  \Phi_{0}\right)  ,
\end{align}
where the value of $\Psi\left(  0\right)  $ depends on the parameter $\Phi
_{0}$ because of $\Phi$ must vanish when $r=1$. This situation can be overcome
redefining the problem as follows:%

\begin{equation}
\Psi\left(  r\right)  =\psi\left(  \xi\right)  +2\ln R_{m},\text{ }\Phi\left(
r\right)  =\varphi\left(  \xi\right)
\end{equation}
being $\left\{  \varphi\left(  \xi\right)  ,\psi\left(  \xi\right)  \right\}
$ the solution of (\ref{structure equation}) whose boundary conditions are
given by:%

\begin{equation}
\varphi\left(  0\right)  =\Phi_{0}>0,\varphi^{\prime}\left(  0\right)
=0,~\psi\left(  0\right)  =0,\psi^{\prime}\left(  0\right)  =0,
\end{equation}
where $\xi$ is related with $r$ throughout the relation:%

\begin{equation}
r=\xi/R_{m},
\end{equation}
being $R_{m}$ the radio at which $\varphi$ vanishes ($\varphi\left(
R_{m}\right)  =0$).

The canonical parameter $\beta$ and $\mu$ are obtained from the relations:%

\begin{equation}
R_{m}\partial_{\xi}\varphi\left(  R_{m}\right)  =-\beta=-\beta\left(  \Phi
_{0}\right)  ,
\end{equation}

\begin{equation}
\psi\left(  R_{m}\right)  +R_{m}\partial_{\xi}\psi\left(  R_{m}\right)  =-h,
\end{equation}
being $h=h\left(  \Phi_{0}\right)  \equiv-\ln\left[  \mathcal{K}\left(
\beta,\mu\right)  /R_{m}^{2}\right]  $, which allow us to express $\mu$ as a
function of $\Phi_{0}$:%

\begin{equation}
\mu=h+\beta-\frac{1}{2}\ln\beta-2\ln R_{m}.
\end{equation}

Taking into consideration the equations (\ref{energy tot}) and
(\ref{planck FP}), the total energy and the Planck potential per particle are
rewritten as follows:%

\begin{equation}
\epsilon\left(  \Phi_{0}\right)  =\frac{3}{2\beta}-\frac{1}{2}-\frac{1}%
{\beta^{2}R_{m}}h_{1}\left(  \Phi_{0}\right)  ,
\end{equation}

\begin{equation}
\mathcal{P}\left(  \Phi_{0}\right)  =-1-\mu+\frac{1}{2}\beta+h\left(  \Phi
_{0}\right)  +\frac{1}{\beta R_{m}}h_{2}\left(  \Phi_{0}\right)  .
\end{equation}
being%

\begin{equation}
h_{1}\left(  \Phi_{0}\right)  =\int_{0}^{R_{m}}4\pi\xi^{2}d\xi\mathbf{~}%
\frac{1}{2}\varphi\left[  F_{1}\left(  \psi,\varphi\right)  +F_{2}\left(
\psi,\varphi\right)  \right]  ,
\end{equation}

\begin{equation}
h_{2}\left(  \Phi_{0}\right)  =\int_{0}^{R_{m}}4\pi\xi^{2}d\xi\text{ }%
F_{1}\left(  \psi,\varphi\right)  \left(  \frac{1}{2}\varphi+\psi\right)  .
\end{equation}

\section{Results and discussions}%

%TCIMACRO{\FRAME{fhFU}{2.597in}{2.4613in}{0pt}{\Qcb{The alternative model
%exhibits the main features of the isothermal model of Antonov. }%
%}{\Qlb{caloric}}{caloric.eps}{\special{ language "Scientific Word";
%type "GRAPHIC";  display "USEDEF";  valid_file "F";  width 2.597in;
%height 2.4613in;  depth 0pt;  original-width 6.2405in;
%original-height 9.0702in;  cropleft "0";  croptop "1";  cropright "1";
%cropbottom "0";  filename 'caloric.eps';file-properties "XNPEU";}}}%
%BeginExpansion
\begin{figure}
[h]
\begin{center}
\includegraphics[
height=2.4613in,
width=2.597in
]%
{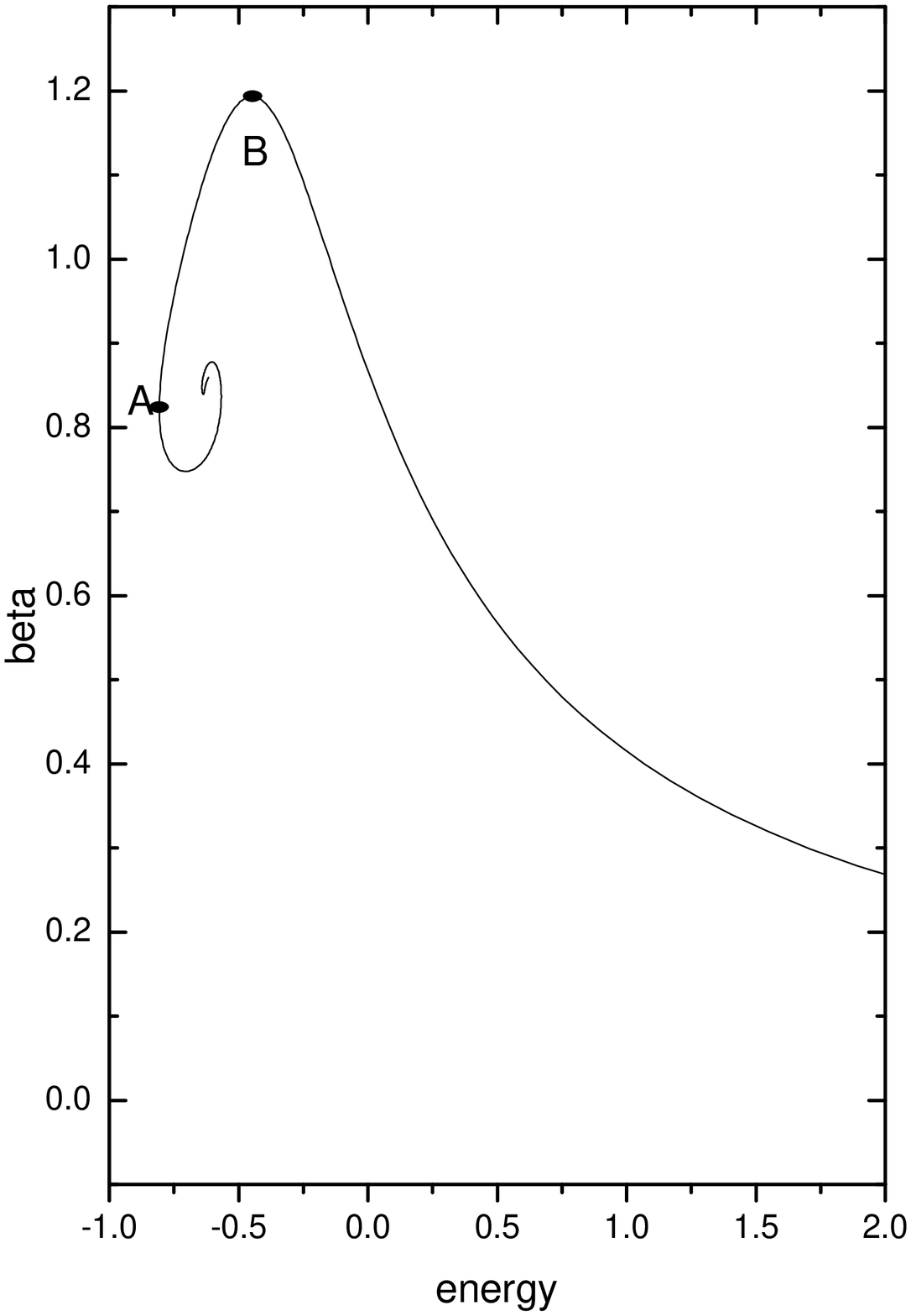}%
\caption{The alternative model exhibits the main features of the isothermal
model of Antonov. }%
\label{caloric}%
\end{center}
\end{figure}
%EndExpansion
%

%TCIMACRO{\FRAME{fhFU}{2.6039in}{2.4613in}{0pt}{\Qcb{Central density
%\textit{versus} energy: this dependence shows the existence of the
%gravitational collapse for $\epsilon<\epsilon_{A}$.}}{\Qlb{density}%
%}{density.eps}{\special{ language "Scientific Word";  type "GRAPHIC";
%display "USEDEF";  valid_file "F";  width 2.6039in;  height 2.4613in;
%depth 0pt;  original-width 6.1713in;  original-height 9.1056in;
%cropleft "0";  croptop "1";  cropright "1";  cropbottom "0";
%filename 'density.eps';file-properties "XNPEU";}}}%
%BeginExpansion
\begin{figure}
[h]
\begin{center}
\includegraphics[
height=2.4613in,
width=2.6039in
]%
{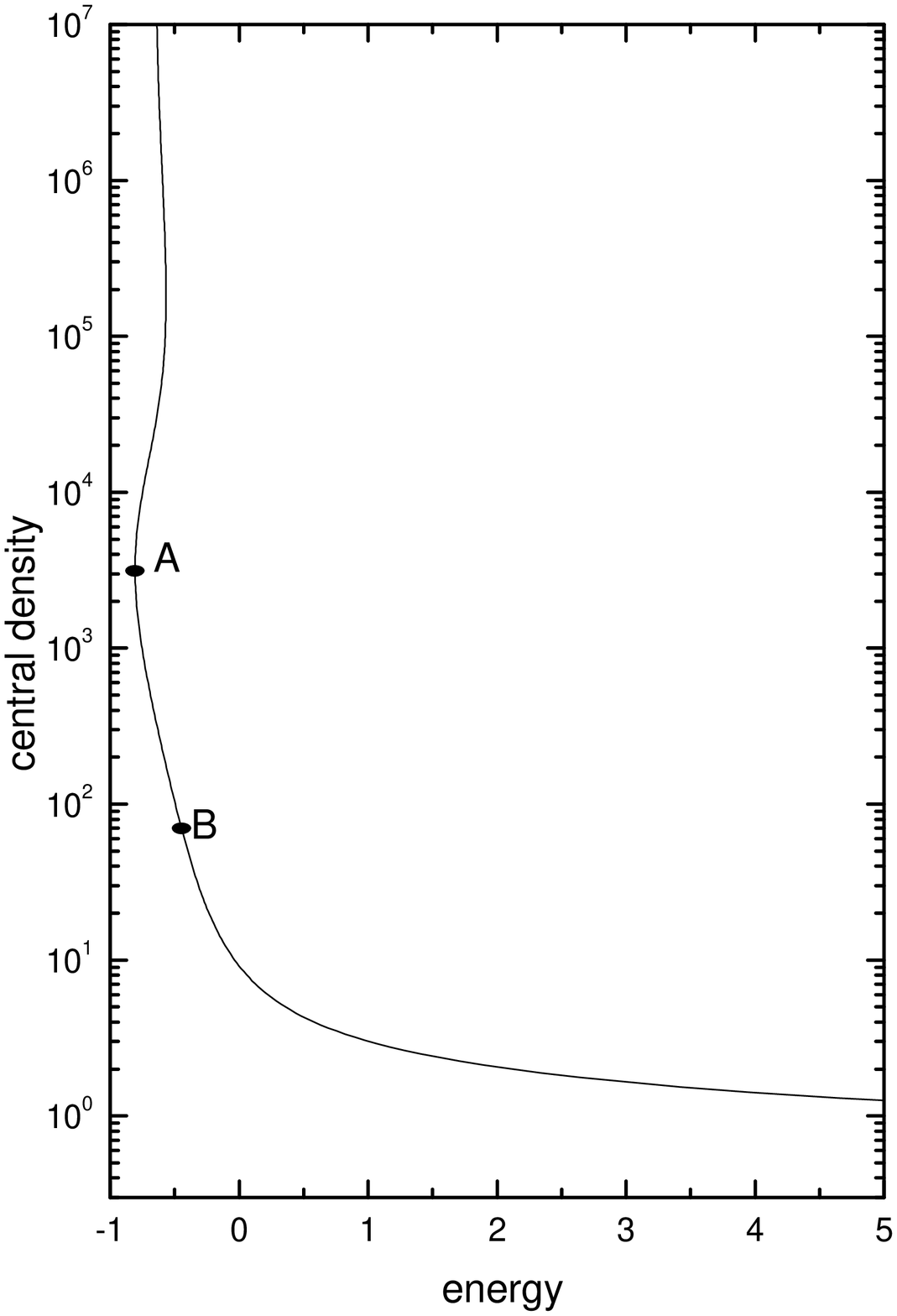}%
\caption{Central density \textit{versus} energy: this dependence shows the
existence of the gravitational collapse for $\epsilon<\epsilon_{A}$.}%
\label{density}%
\end{center}
\end{figure}
%EndExpansion

Figures (\ref{caloric}) and (\ref{density}) show respectively the caloric
curve and the central density of the model. These dependences evidence that
this model system exhibits the main features of the Antonov isothermal model:
the existence of a negative specific heat when $\epsilon_{A}<\epsilon
<\epsilon_{B}$, and the gravitational collapse for $\epsilon<\epsilon_{A}$,
where\ the central density grows towards the infinity and the system develops
a core-halo structure, being $\epsilon_{A}=-0.806$ and $\epsilon_{B}=-0.446$.

\begin{figure}
[h]
\begin{center}
\includegraphics[
height=2.4613in,
width=2.6282in
]%
{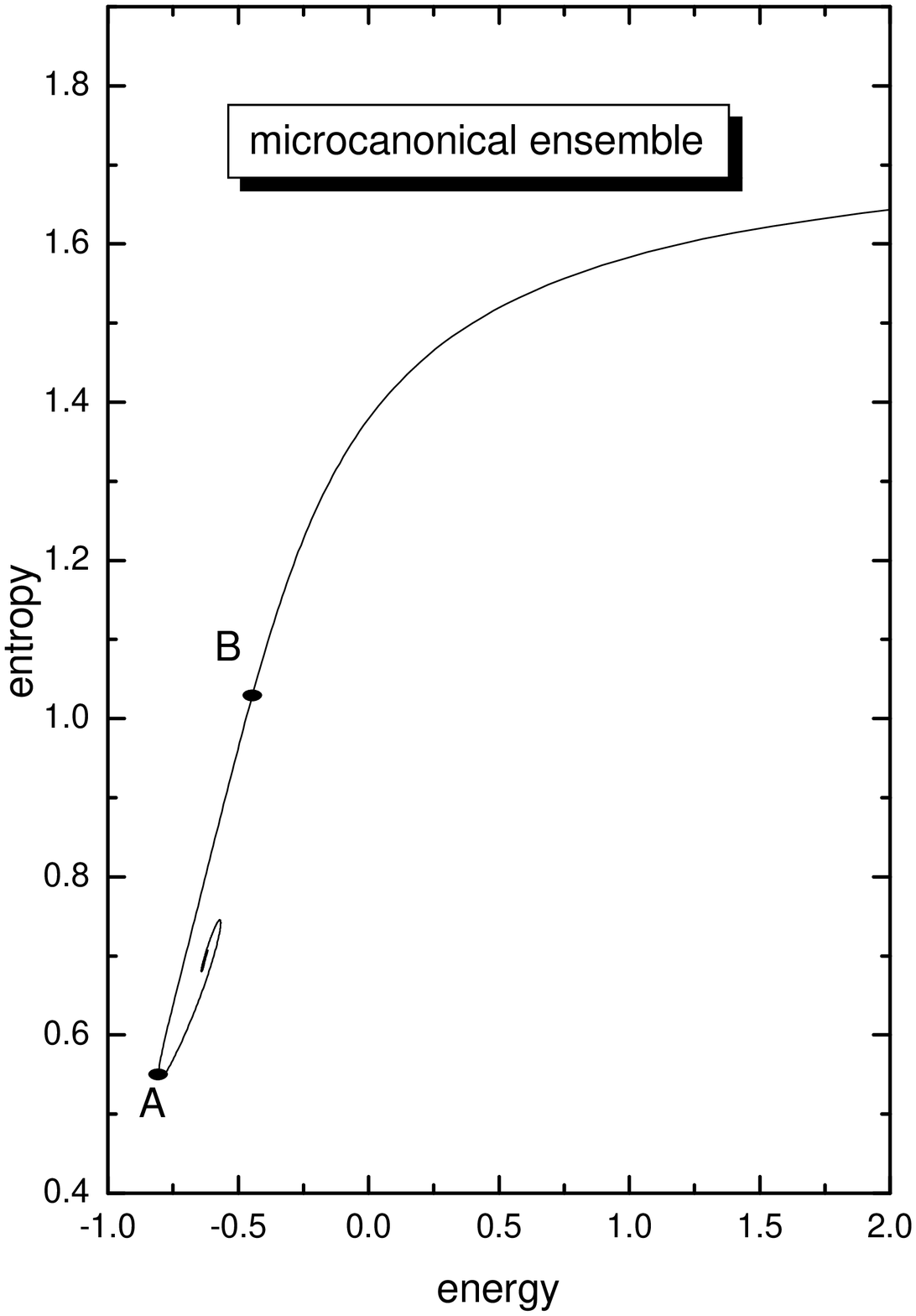}%
\caption{Entropy versus energy: Equilibrium configurations exit only for
$\epsilon>\epsilon_{A}$. }%
\label{micro}%
\end{center}
\end{figure}

\begin{figure}
[h]
\begin{center}
\includegraphics[
height=2.4613in,
width=2.6091in
]%
{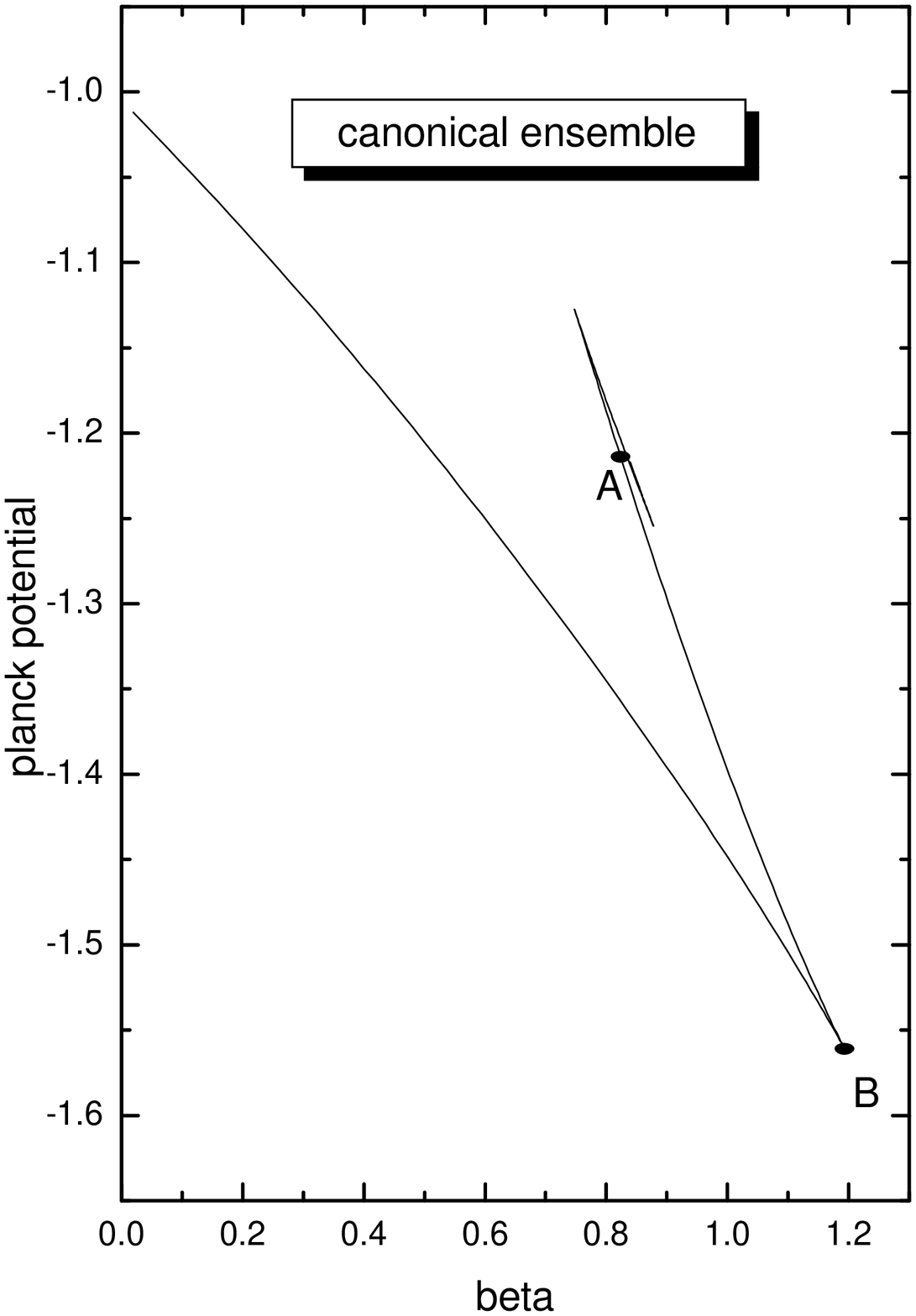}%
\caption{Planck potential \textit{versus} inverse temperature: the canonical
ensemble access only to those equilibrium stages with $\beta<\beta_{B}$
($\epsilon>\epsilon_{B}$).}%
\label{canon}%
\end{center}
\end{figure}

This conclusion is supported by the analysis of the thermodynamical potentials
of the model: the entropy in the microcanonical ensemble, at figure
(\ref{micro}), and the Planck potential in the canonical ensemble, at figure
(\ref{canon}). The figure (\ref{micro}) shows that the points of the superior
branch of the caloric curve (\ref{caloric}) correspond to equilibrium
configuration while the others represent unstable saddle points. No
equilibrium states exist when $\epsilon<\epsilon_{A}$.

On the other hand, the figure (\ref{canon}) evidences that the canonical
ensemble can access only to those equilibrium states beloging to the interval
$0<\beta<\beta_{B}$ (with $\epsilon>\epsilon_{B}$), where $\beta_{B}=1.193$.
The energetic region $\epsilon_{A}<\epsilon<\epsilon_{B}$ is invisible for the
canonical description due to the negativity of the heat capacity. No
equilibrium states exist for $\beta>\beta_{B}$. This fact evidences the
existence of a gravitational collapse in the canonical ensemble beyond the
critical point $\beta_{B}$, which is usually refered as an \textit{isothermal
collapse} \cite{chava3}. %

%TCIMACRO{\FRAME{fhFU}{2.6282in}{2.4613in}{0pt}{\Qcb{Entropy versus energy:
%Equilibrium configurations exit only for $\epsilon>\epsilon_{A}$. }%
%}{\Qlb{micro}}{micro.eps}{\special{ language "Scientific Word";
%type "GRAPHIC";  display "USEDEF";  valid_file "F";  width 2.6282in;
%height 2.4613in;  depth 0pt;  original-width 6.2405in;
%original-height 9.0148in;  cropleft "0";  croptop "1";  cropright "1";
%cropbottom "0";  filename 'micro.eps';file-properties "XNPEU";}}}%
%BeginExpansion
%EndExpansion
%

%TCIMACRO{\FRAME{fhFU}{2.6091in}{2.4613in}{0pt}{\Qcb{Planck potential
%\textit{versus} inverse temperature: the canonical ensemble access only to
%those equilibrium stages with $\beta<\beta_{B}$ ($\epsilon>\epsilon_{B}$).}%
%}{\Qlb{canon}}{canon.eps}{\special{ language "Scientific Word";
%type "GRAPHIC";  display "USEDEF";  valid_file "F";  width 2.6091in;
%height 2.4613in;  depth 0pt;  original-width 6.1981in;
%original-height 8.9448in;  cropleft "0";  croptop "1";  cropright "1";
%cropbottom "0";  filename 'canon.eps';file-properties "XNPEU";}}}%
%BeginExpansion
%EndExpansion

The Gibbs' argument, the equilibrium of a subsystem with a thermal bath, is
non applicable to this situation because of no reasonable thermal bath exits
for the astrophysical systems.Therefore, the isothermal catastrophe is not a
phenomenon with physical relevance since it can be never obtained in nature:
the consideration of a thermal bath in the astrophysical system is outside of
context. A different significance possesses the gravothermal catastrophe. The
gravitational collapse is the main engine of structuration in astrophysics and
it concerns almost all scales of the universe: the formation of planetesimals
in the solar nebula, the formation of stars, the fractal nature of the
interstellar medium, the evolution of globular clusters and galaxies and the
formation of galactic clusters in cosmology \cite{chava3}.

Let us now carry out a comparative study between the isothermal model and this
alternative onel. As already mentioned, the only difference between these
approaches relies on the regularization prescription of the long-range
singularity of the selfgravitating gas: the isothermal model avoids the
particles evaporation by using a rigid container, while the present model
takes into account the effect of this evaporation by truncating the kinetic
energy distribution function. This study is carried out by considering a
spherical container in the isothermal model with a linear dimension
$R=R_{tidal}$.%

%TCIMACRO{\FRAME{fhFU}{2.5953in}{2.4613in}{0pt}{\Qcb{The caloric curves.The
%figure shows the comparison between the two models.}}{\Qlb{com_cal}%
%}{com_cal.eps}{\special{ language "Scientific Word";  type "GRAPHIC";
%display "USEDEF";  valid_file "F";  width 2.5953in;  height 2.4613in;
%depth 0pt;  original-width 6.1713in;  original-height 9.0373in;
%cropleft "0";  croptop "1";  cropright "1";  cropbottom "0";
%filename 'com_cal.eps';file-properties "XNPEU";}}}%
%BeginExpansion
\begin{figure}
[h]
\begin{center}
\includegraphics[
height=2.4613in,
width=2.5953in
]%
{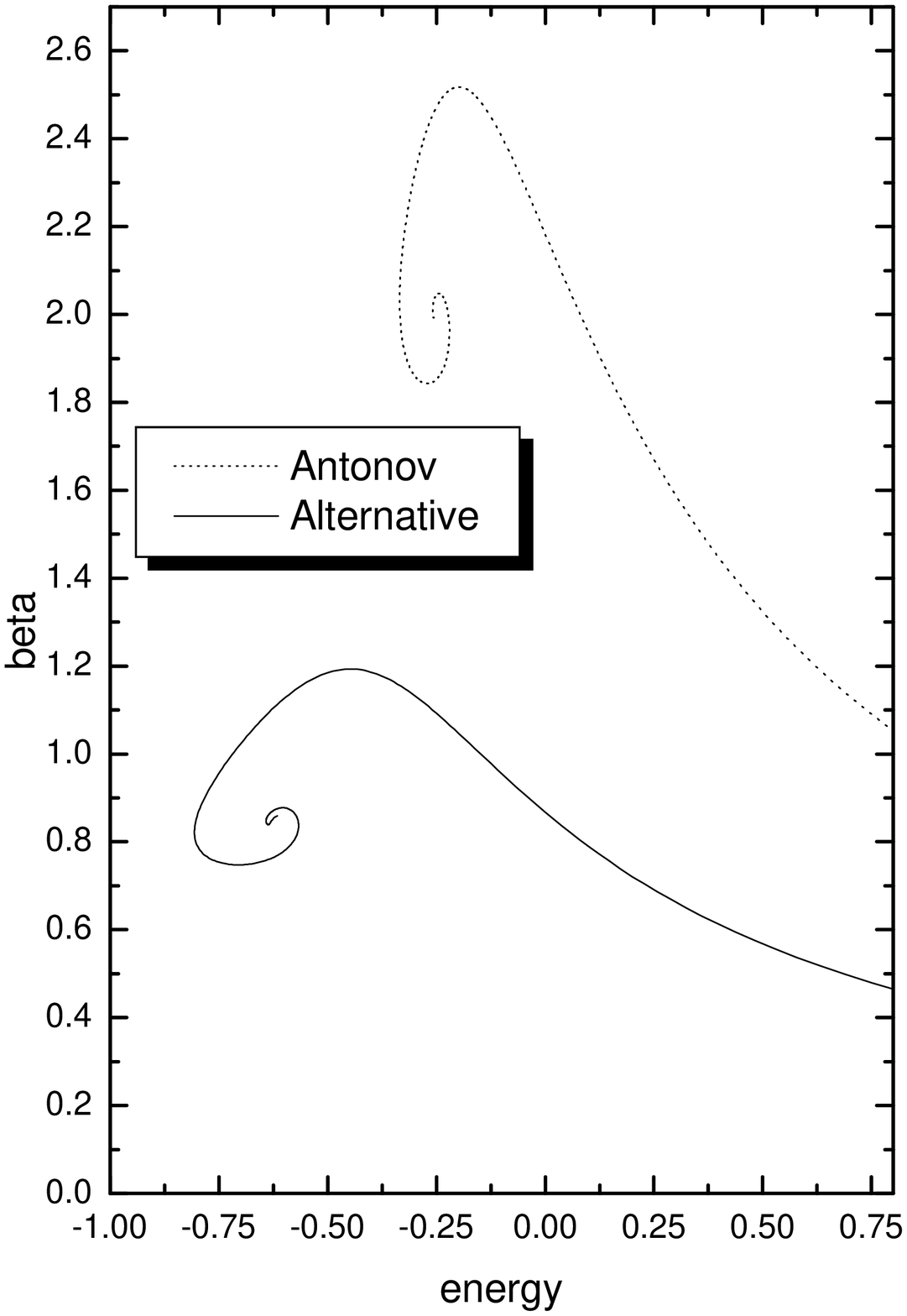}%
\caption{The caloric curves.The figure shows the comparison between the two
models.}%
\label{com_cal}%
\end{center}
\end{figure}
%EndExpansion
%

%TCIMACRO{\FRAME{fhFU}{2.6792in}{2.4613in}{0pt}{\Qcb{The central density
%\textit{versus} energy. The central density is much greater in this
%alternative version of the Antonov problem.}}{\Qlb{com_den}}{com_den.eps}%
%{\special{ language "Scientific Word";  type "GRAPHIC";  display "USEDEF";
%valid_file "F";  width 2.6792in;  height 2.4613in;  depth 0pt;
%original-width 6.6997in;  original-height 9.0702in;  cropleft "0";
%filename 'com_den.eps';file-properties "XNPEU";}}}%
%BeginExpansion
\begin{figure}
[h]
\begin{center}
\includegraphics[
height=2.4613in,
width=2.6792in
]%
{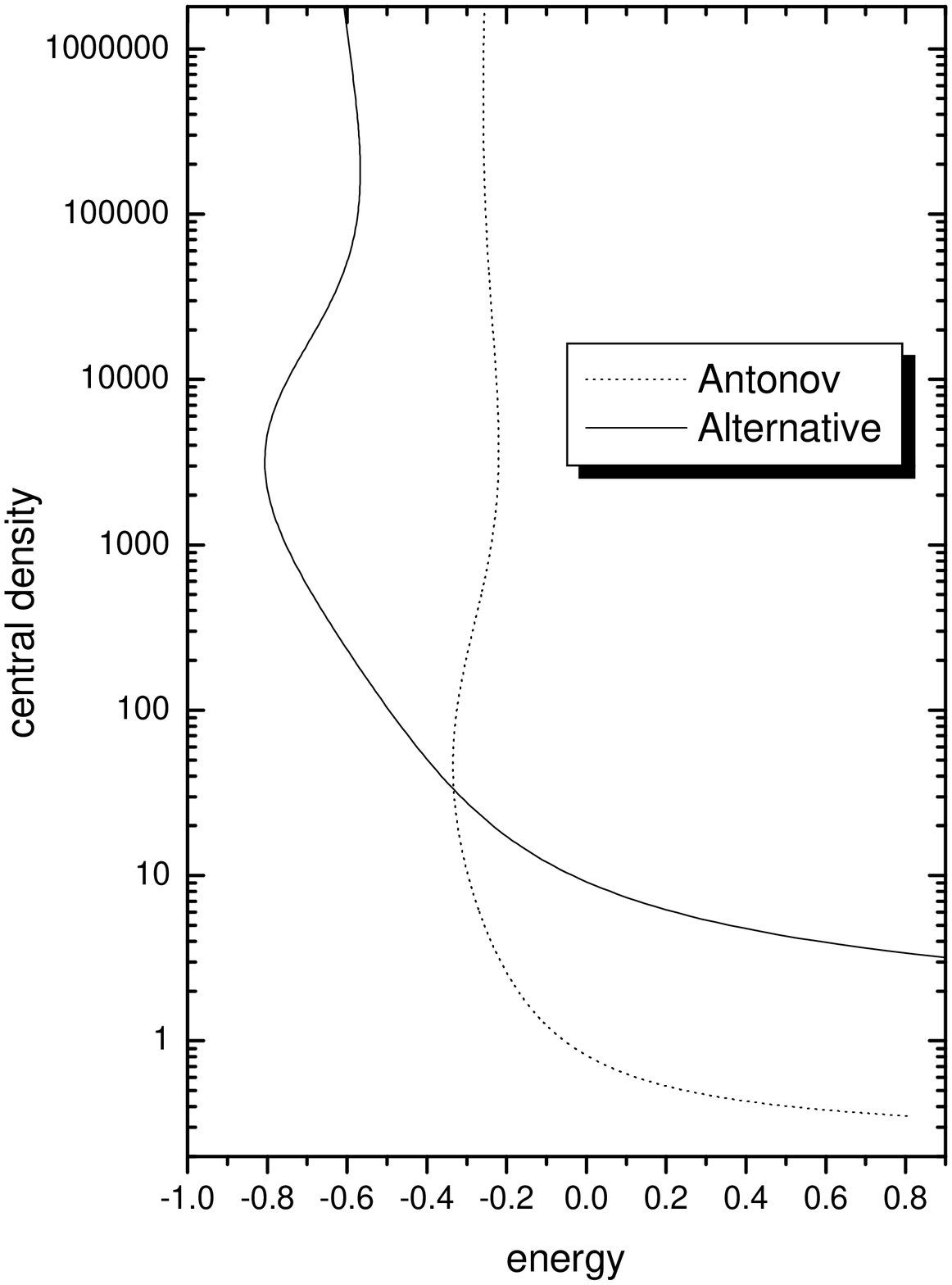}%
\caption{The central density \textit{versus} energy. The central density is
much greater in this alternative version of the Antonov problem.}%
\label{com_den}%
\end{center}
\end{figure}
%EndExpansion

Figure (\ref{com_cal}) shows the caloric curves of these models. In spite of
the qualitative similarity of these dependences, the alternative model is able
to describe an additional energetic range: from $-0.806$ to $-0.335$; but the
isothermal model describes equilibrium configurations belonging to the
interval $1.193<\beta<2.518$, which are cooler than the ones described by
using the first model..

A second difference is evidenced in regard to the central density
\textit{versus} energy dependence, which is shown in the figure (\ref{com_den}%
). The central density at the critical energy of the gravitational collapse is
much greater in the alternative model than the isothermal one, and this
qualitative relationship seems to be applicable to the whole energetic range.%

%TCIMACRO{\FRAME{fhFU}{2.5867in}{2.4613in}{0pt}{\Qcb{Isothermal core radio and
%its mass content in the alternative model.}}{\Qlb{core}}{core.eps}%
%{\special{ language "Scientific Word";  type "GRAPHIC";  display "USEDEF";
%valid_file "F";  width 2.5867in;  height 2.4613in;  depth 0pt;
%original-width 6.0597in;  original-height 9.0926in;  cropleft "0";
%filename 'core.eps';file-properties "XNPEU";}}}%
%BeginExpansion
\begin{figure}
[h]
\begin{center}
\includegraphics[
height=2.4613in,
width=2.5867in
]%
{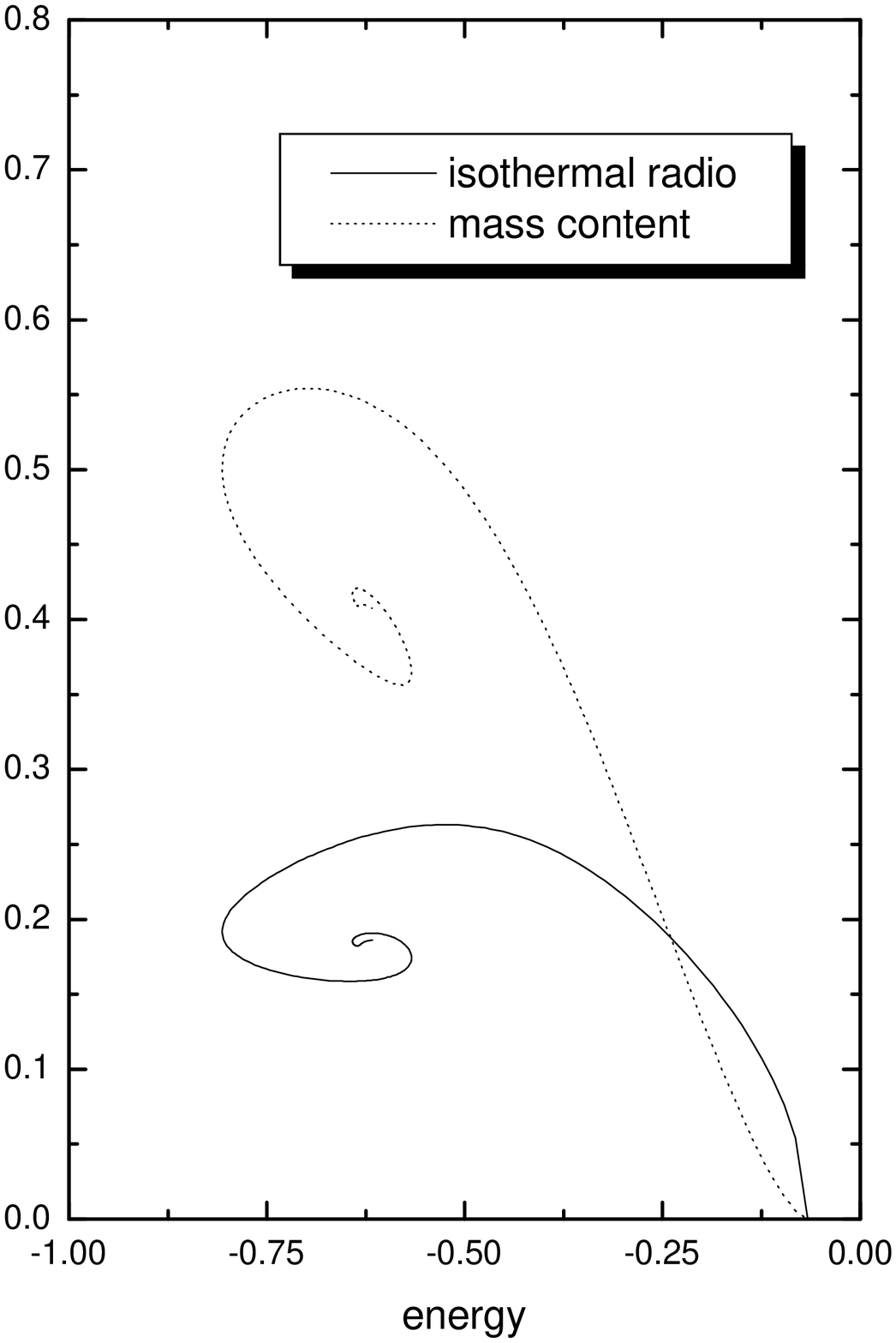}%
\caption{Isothermal core radio and its mass content in the alternative model.}%
\label{core}%
\end{center}
\end{figure}
%EndExpansion

The figure (\ref{core}) shows two interesting observables: the radio in which
the system exhibits an isothermic behavior, $R_{i.c}$, and the mass content
enclosed inside, $M_{i.c}$. This quantities characterizes the size and the
mass of the isothermal core. The reader may observe that the isothermal core
contains at the critical energy of the gravitational collapse, the half of the
system total mass inside the $\sim1\%$ \ of the system volume. On the other
hand, it is interesting to note that this isothermal core disappears at
$\epsilon^{\ast}\simeq-0.066$.

The quantitative differences in the thermodynamical description between these
models seem to be explained by the following reasonings. Most of the energy
contribution to the total energy comes from the isothermal core. The
contribution of the gravitational potential energy is dominant in the core due
to the high mass concentration enclosed inside this region, where moreover,
the Newtonian potential exhibits its highest values. Outside the isothermal
core, the kinectic energy contribution decreases considerably as consequence
of the deviation from the isothermal character of the microscopic particles
distribution function due to the evaporation.

These arguments can be rephrased as follows: in the isothermal region, where
$-0.806<\epsilon<-0.066$, the present model behaves as an isothermal model
with a characteristic size equal to the size of the isothermal core. Since the
relevant canonical variables in the isothermal model is $\eta=\beta GM/R$
\cite{antonov,chava3}, a rough estimation of the maximal value $\beta_{\max
}=\beta_{B}$ in which the isothermal collapse takes places is given by:%

\begin{equation}
\beta_{B}\approx\frac{R_{i.c}}{M_{i.c}}\beta_{\max}^{isoth.}%
.\label{estimation}%
\end{equation}
By using the characteristics values $R_{i.c}\approx0.2$, $M_{i.c}\approx0.5$,
and $\beta_{\max}^{isoth.}\approx2.5$ (obtained from the isothermal model),
the formula (\ref{estimation}) gives a fairly good estimation of $\beta
_{B}\approx1$ ($\beta_{B}=1.193$).

The main difference between these model is in regard to the character of the
equilibrium profiles in the high energy region with $\epsilon>0$. As already
discussed, the isothermal core has disappear in this energetic region and the
equilibrium configurations of the system are essentially polytropic, while the
isothermal model leads to a uniform distribution of the particles throughout
the volume of the rigid container.

In this case, the structure equations (\ref{structure equation}) become in a
\textit{quasi-polytropic model}:%

\begin{align}
\Delta\Phi &  =-4\pi\exp\left(  \Psi\right)  \frac{4}{3\sqrt{\pi}}\Phi
^{\frac{3}{2}},\text{ \ }\nonumber\\
\Delta\Psi &  =-4\pi\exp\left(  \Psi\right)  \frac{2}{\sqrt{\pi}}\Phi
^{\frac{1}{2}},\label{pomod}%
\end{align}
which exhibits the same fractal characteristics of a polytropic model with
polytropic index $\gamma=\frac{5}{3}$. This polytropic index characterizes an
adiabatic process of the ideal gas of particles, which is in our case the
evaporation of the system in the vacuum. However, this equation system is not
equivalent to the polytropic model due to the presence of the driving function
$C$. This purely polytropic profile is obtained by disregarding the second
equation and setting $\Psi\equiv0$ in the first one.%

%TCIMACRO{\FRAME{fhFU}{2.6567in}{2.4535in}{0pt}{\Qcb{Comparison among the
%equilibrium profiles: A) Alternative model with an isothermal core, B)
%Isothermal profile, C) Alternative model with a quasi-polytropic profile, and
%D) Purely polytropic profile.}}{\Qlb{profiles}}{com_prof.eps}%
%{\special{ language "Scientific Word";  type "GRAPHIC";  display "USEDEF";
%valid_file "F";  width 2.6567in;  height 2.4535in;  depth 0pt;
%original-width 6.5881in;  original-height 9.1099in;  cropleft "0";
%filename 'com_prof.eps';file-properties "XNPEU";}}}%
%BeginExpansion
\begin{figure}
[h]
\begin{center}
\includegraphics[
height=2.4535in,
width=2.6567in
]%
{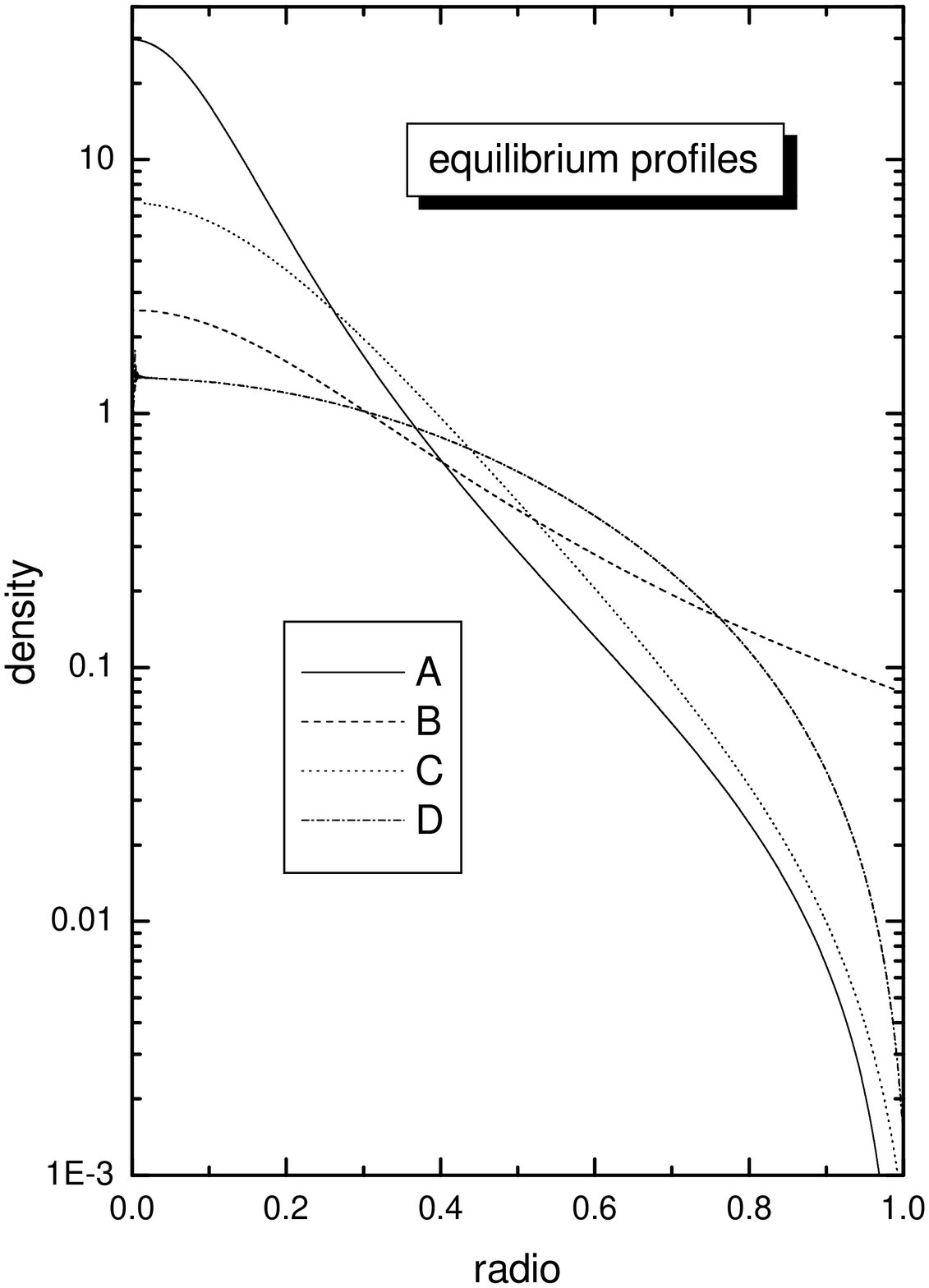}%
\caption{Comparison among the equilibrium profiles: A) Alternative model with
an isothermal core, B) Isothermal profile, C) Alternative model with a
quasi-polytropic profile, and D) Purely polytropic profile.}%
\label{profiles}%
\end{center}
\end{figure}
%EndExpansion

Let us finalize our discussion by carrying out a comparative study among the
equilibrium profiles obtained from the alternative model with the well-known
isothermal and polytropic profiles. These equilibrium profiles are shown in
the figure (\ref{profiles}). Equilibrium profiles A and C were obtained from
the alternative model: profile A corresponds to an equilibrium configuration
possessing an isothermal core, while C is a quasi-polytropic equilibrium
profile. Profiles B and D correspond respectively to an isothermal and
polytropic configurations. The reader can note that the alternative model
profiles differ essentially in regard to the existence of the isothermal core,
since no qualitative differences are evidenced in the halo structure. On the
other hand, these results suggest that a system undergoing an evaporation
process concentrates their particles in the inner regions more than what the
isothermal or the polytropic models predict.

\section{Conclusions}

As already shown in the present paper, the consideration of the energetic
prescription (\ref{renormalization}) leads to an alternative version of the
Antonov problem. This model, besides of exhibiting the main features of the
isothermal model: the gravitational collapse and the energetic region with a
negative specific heat, has the advantage that the system size naturally
appears as consequence of the particles evaporation. There is no need of
enclosing the system in a rigid container in order to avoid long-range
singularity of the Newtonian potential because this regularization procedure
is sufficient to access to finite equilibrium configurations.

It is remarkable that the present approach unifies the well-known isothermal
and polytropic equilibrium profiles. As already mentioned, the equilibrium
profiles derived from this alternative model differ essentially in regard to
the existence of the isothermal core, since no qualitative differences are
evidenced in the halo structure. The comparative study of the equilibrium
$\rho$ profiles suggests that a system undergoing an evaporation process
concentrates the particles in the inner regions more than what predict by the
isothermal or the polytropic models.

There are many open questions in the study of this model system by using this
kind of regularization scheme, as example, the dynamical aspects: Which is the
influence of the particles evaporation in the system dynamical evolution? and
how could it be performed this dynamical description by preserving this
energetic prescription? Further studies should clarify these questions. The
present study is carried out by considering that the evaporation rate is small
enough in order to ensure that the system reaches a quasistationary state.
This assumption is satisfied by the globular clusters, since the relaxation
time $t_{relax}$ in them differs considerable from the evapotation time
$t_{evap}$, $t_{relaz}\ll t_{evap}$ (see in ref.\cite{berna}). The dynamical
approch could be developed by considering that the system evolutes slowly
throughout these quasistationary states.

\end{document}